# Performance limits and benchmarking of ideal 2D topological insulator field-effect transistors

**Yosep Park[1], Jaeyoung Eom[2], and Yeonghun Lee[1, 2, 3, *]**

***Abstract***: Topological insulator field-effect transistors (TIFETs) have emerged as a promising alternative to overcome the scaling limits of silicon-based field-effect transistors (Si-based FETs). In this study, we establish the performance limits of ideal TIFETs and present a comprehensive benchmarking of the switching energy-delay performance of TIFETs against Si-based FETs (FinFET and LGAA) and 2-D material-based FETs ($MoS_2$ and InSe) under equivalent and realistic conditions. Furthermore, we analyze the key factors influencing the switching energy-delay performance of TIFETs, including contact resistance ($R_C$), parasitic capacitance ($C_{par}$), channel length ($L$), and channel width ($W$). Notably, TIFETs achieve substantially lower switching energy and delay time, outperforming both Si-based FETs and 2-D material-based FETs, where a transition from 1-stack to 3-stack yields approximately a sixfold reduction. This benchmarking study shows that TIFETs exhibit competitive switching performance and hold strong potential for future topological electronic devices.

***Index terms***: Topological transistors, topological insulators, benchmarking, switching energy, delay

## I. INTRODUCTION

The International Roadmap for Devices and Systems (IRDS) predicted the physical limits of scaling and supply voltage reduction of Si-based FETs [1-4]. To address these limitations, a variety of next-generation devices employing alternative switching mechanisms have been extensively investigated [5-8]. Among them, TIFETs have emerged as a promising candidate [9-15]. Two-dimensional topological insulators (2D TIs), the core material in TIFETs, offer two key advantages. First, they exhibit an electrically tunable quantum phase transition, where the quantum spin Hall (QSH) phase is transformed into the trivial insulator phase by a broken inversion symmetry [9, 16, 17]. Second, 2D TIs have helical edge states that exhibit ballistic transport protected by time-reversal symmetry, which prevents backscattering. [16,17]

[1]Department of Intelligent Semiconductor Engineering,
Incheon National University, Incheon 22012, Republic of Korea.
[2]Department of Electronics Engineering,
Incheon National University, Incheon 22012, Republic of Korea
[3]Research Institute for Engineering and Technology,
Incheon National University, Incheon 22012, Republic of Korea
*E-mail: y.lee@inu.ac.kr (Corresponding Author: Yeonghun Lee)

These distinctive properties of topological insulators—phase-transition-driven switching and dissipationless edge transport—make TIFETs as compelling alternatives to conventional FETs.

The two aforementioned advantages of 2D TIs have motivated investigations into the switching energy-delay performance of TIFETs. For instance, prior work has explored TIFETs based on stanene/$SiO_2$ layers, employing non-equilibrium Green's function (NEGF) modeling for transport characteristics and analyzing energy-delay performance by incorporating quantum capacitance into the gate capacitance model [18]. Similarly, TIFETs utilizing monolayer $Na_3Bi/SiO_2$ have shown promising energy-delay performance at low drain voltages [19]. Recent studies on bismuthene/h-BN nanoribbon TIFETs have demonstrated the potential for enhanced energy-delay performance, indicating superiority over the IRDS criterion for a 10nm gate length [15]. However, these prior studies [12, 18-20] have limitations in making direct comparison with other devices because device parameters such as channel width, channel length, equivalent oxide thickness (EOT), total gate capacitance ($C_{g,tot}$), and supply voltage ($V_{DD}$) were not consistently aligned across different devices. Consequently, it has been challenging to accurately quantify the performance advantages of TIFETs in terms of switching energy-delay compared to charge-based devices like IRDS FETs or other 2D material-based FETs. Additionally, the prior benchmarking study [12] was based on a disorder-driven switching mechanism, which is fundamentally distinct from the topological-phase-transition-driven mechanism addressed here.

In this work, we present a comprehensive analysis of the impact of device parameters such as $R_{\mathrm{C}}$, $C_{\mathrm{par}}$, $L$, $W$ on TIFET switching energy-delay performance. Subsequently, we benchmark the switching energy-delay performance of TIFETs against Si-based FETs [1] (FinFET and LGAA) from the IRDS 2023 and 2025 targets for high-performance (HP) logic transistors, as well as representative 2D FETs ($MoS_2$ [21] and InSe [22]). Furthermore, although our current study assumed monolayer $MoS_2$ as the topological material, we anticipate that this work will establish a general TIFET energy/delay reference point, as changes in the specific TI material are not expected to result in a substantial alteration of the overall conclusions regarding the maximum switching energy-delay performance of TIFETs. Therefore, our comprehensive benchmark demonstrates that TIFETs have competitive switching energy and delay time performance compared to Si-based FETs and 2D FETs, while providing valuable insights into the performance limits and future prospects of topological transistors.

## II. Device structures

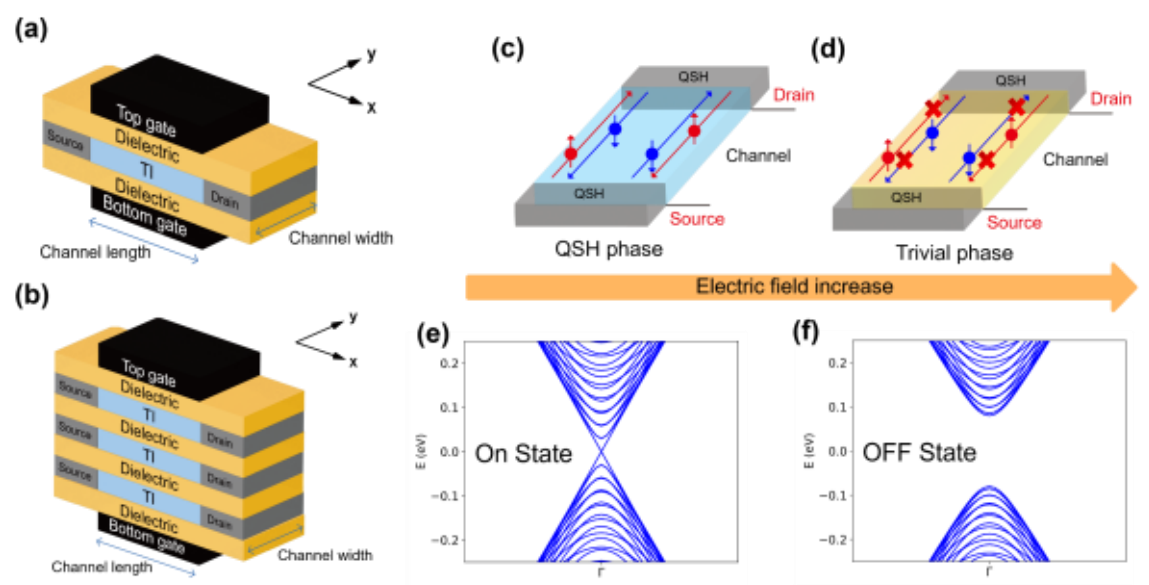

**Fig. 1.** The device structure and topological phase transitions. (a,b) Structure of the 1-stack TIFET and 3-stack TIFET. (c) The QSH phase with ballistic edge transport for the on-state and (d) the trivial phase for the off-state without the edge transport in the out-of-plane electric field ($E_Z$). Band structures of TI nanoribbons in (e) the QSH phase at on-state and (f) the trivial phase at off-state, simulated using the tight-binding model for a representative TI, stanene [9].

As illustrated schematically in Fig. 1 (a-b), TIFETs consist of a double-gate structure with a topological insulator (TI) channel sandwiched by dielectric materials and gates, where top gate voltage ($V_{\mathrm{G}}$) and bottom gate voltage ($V_{\mathrm{B}}$) are applied. Additionally, both the source and drain regions are constructed of the same material as the 2D TI in the channel. This paper mainly proposes two configurations of TIFETs: 1-stack TIFET (monolayer channel) as illustrated in Fig. 1(a) and the 3-stack TIFET (three-layer channel) shown in Fig. 1(b). Fig. 1(c-e) illustrates the topological phase transition of the 2D TIs controlled by $E_Z$. The ON-state is evaluated at $V_G - V_B = 0$ V ($E_Z = 0$), where the QSH phase yields the upper-bound ON-current governed by $G_0$ [9]. In contrast, increasing $|V_G - V_B|$ induces a topological phase transition into the trivial phase, thereby turning off the conduction. Regarding the channel material, a number of 2D TIs have been identified, including silicene [23,24], germanene [23,25], stanene [9,23], 1T′- $MoS_2$ [26,27], and hafnium-based compounds [28,29].

## III. Results and discussion

In the benchmarking analysis, the delay time ($\tau_{\mathrm{delay}}$) of TIFET is calculated as

$$\tau_{\mathrm{delay}} = \frac{C_{\mathrm{g,tot}} \times V_{\mathrm{DD}}}{I_{\mathrm{D}}}, \qquad (1)$$

where $C_{\mathrm{g,tot}}$ defined as the sum of $C_{\mathrm{g}}$ and $C_{\mathrm{par}}$, and $I_{\mathrm{D}}$ is the drain current that includes $R_{\mathrm{C}}$ effects (see Eq. (6)). Because the topological phase transition is driven by the $E_Z$ between the top and bottom gates, the intrinsic gate capacitance ($C_g$) is determined by the series connection of the top oxide, channel, and bottom oxide capacitances. For a 1-stack TIFET, $C_g$ is expressed as Eq. (2):

$$\frac{1}{C_{\mathrm{g}}} = \frac{1}{C_{\mathrm{OX}}} + \frac{1}{C_{\mathrm{channel}}} + \frac{1}{C_{\mathrm{OX}}}. \qquad (2)$$

For the 3-stack TIFET, the intrinsic gate capacitance is expressed as Eq. (3):

$$\frac{1}{C_{\mathrm{g}}} = \frac{4}{C_{\mathrm{OX}}} + \frac{3}{C_{\mathrm{channel}}}. \qquad (3)$$

The oxide capacitance is $C_{\mathrm{OX}} = (\varepsilon_0 \varepsilon_{\mathrm{SiO2}} WL)/\mathrm{EOT}$, and the channel capacitance is $C_{\mathrm{channel}} = (\varepsilon_{\mathrm{ch}} \varepsilon_0 WL)/t_{\mathrm{ch}}$, where $W$ and $L$ are the channel width and length of each channel layer, respectively, $\varepsilon_0$ is vacuum permittivity, $\varepsilon_{ch}$ is the channel dielectric constant, and $\varepsilon_{SiO2}$ is the relative permittivity of silicon dioxide.

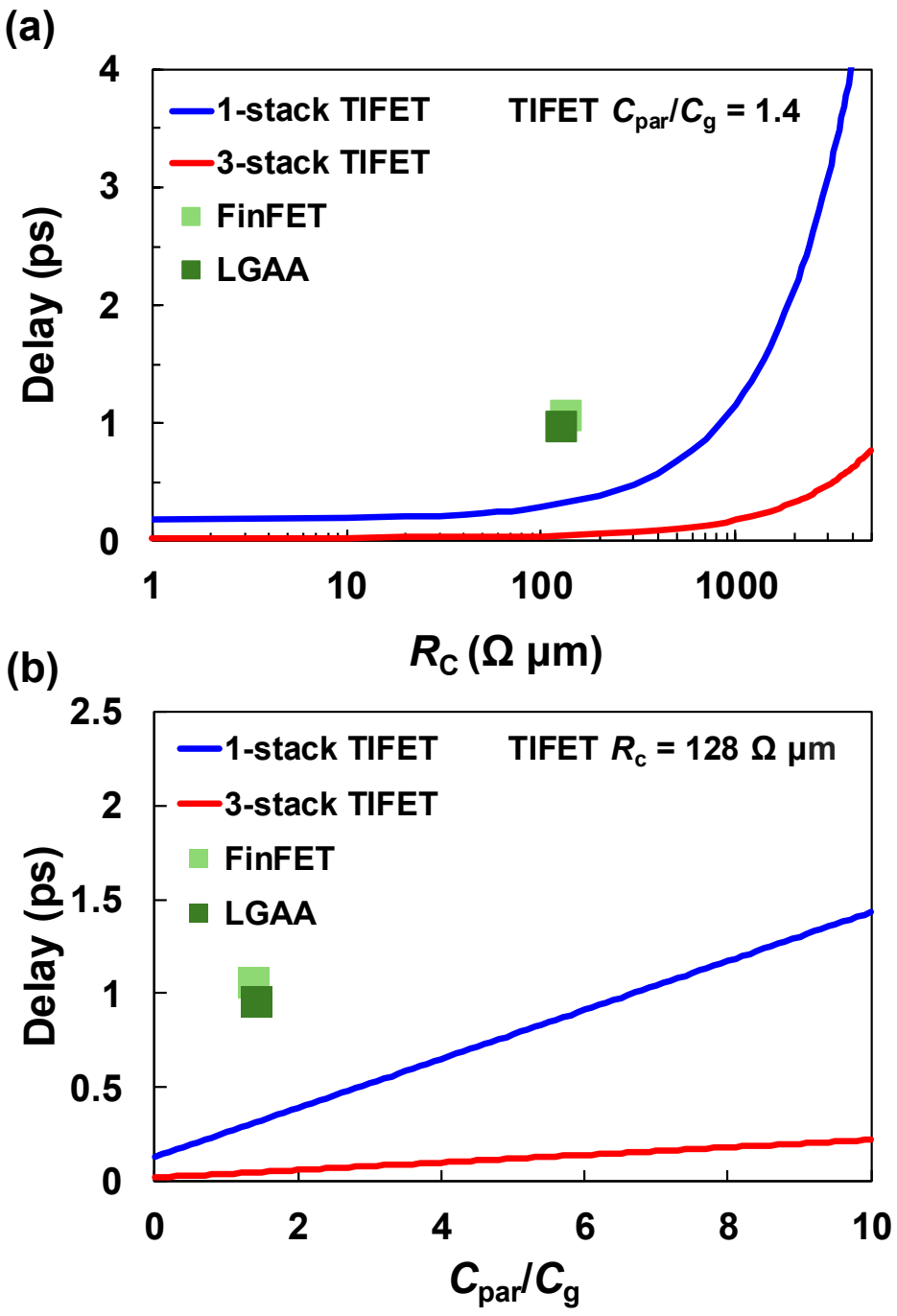

**Fig. 2.** Impact of $R_C$ and $C_{par}$ on the delay time of TIFETs. (a) Delay time as a function of $R_C$ with fixed parasitic capacitance to intrinsic gate capacitance ratio ($C_{par}/C_g$) of 1.4. (b) Delay time as a function of the $C_{par}/C_g$ with fixed $R_C$ = 128 Ω·μm. The data for FinFET and LGAA devices were obtained from the IRDS 2023 and 2025 targets for HP logic transistors, respectively. Note that the $R_C$ value used for FinFET and LGAA represents half of the total parasitic series resistance ($R_{SD}$) of 271 Ω·μm and 257 Ω·μm as reported in the IRDS [1].

**Table 1.** Dielectric constants and monolayer thicknesses of various 2D materials.

| | Dielectric constant | Monolayer thickness (nm) |
|---|---|---|
| 2H-$MoS_2$ [30] | 6.2 | 0.611 |
| 2H-$MoSe_2$ [30] | 7.2 | 0.649 |
| 2H-$MoTe_2$ [30] | 9.1 | 0.71 |
| 2H-$WS_2$ [30] | 6.1 | 0.612 |
| 2H-$WSe_2$ [30] | 7.4 | 0.648 |
| h-BN [30] | 3.29 | 0.317 |
| $HfS_2$ [30] | 6.6 | 0.574 |
| $HfSe_2$ [30] | 7.6 | 0.613 |
| Stanene | 1.6 [36] | 0.45 [37] |
| Silicene | 1.47 [36] | 0.6 [38] |
| Germanene | 1.51 [36] | 0.6 [39] |
| Graphene | 6.9 [40] | 0.26 [41] |

To accurately evaluate the delay time, it is necessary to account for the influence of $R_C$ on the drain current. Based on the ballistic edge transport characteristics of topological insulators, the ideal drain current ($I_D{}'$) without $R_C$ is given by

$$I_D' = G_0 \times V_{DD}, \quad (4)$$

where $G_0 = 2e^2/h$ is the quantum conductance.

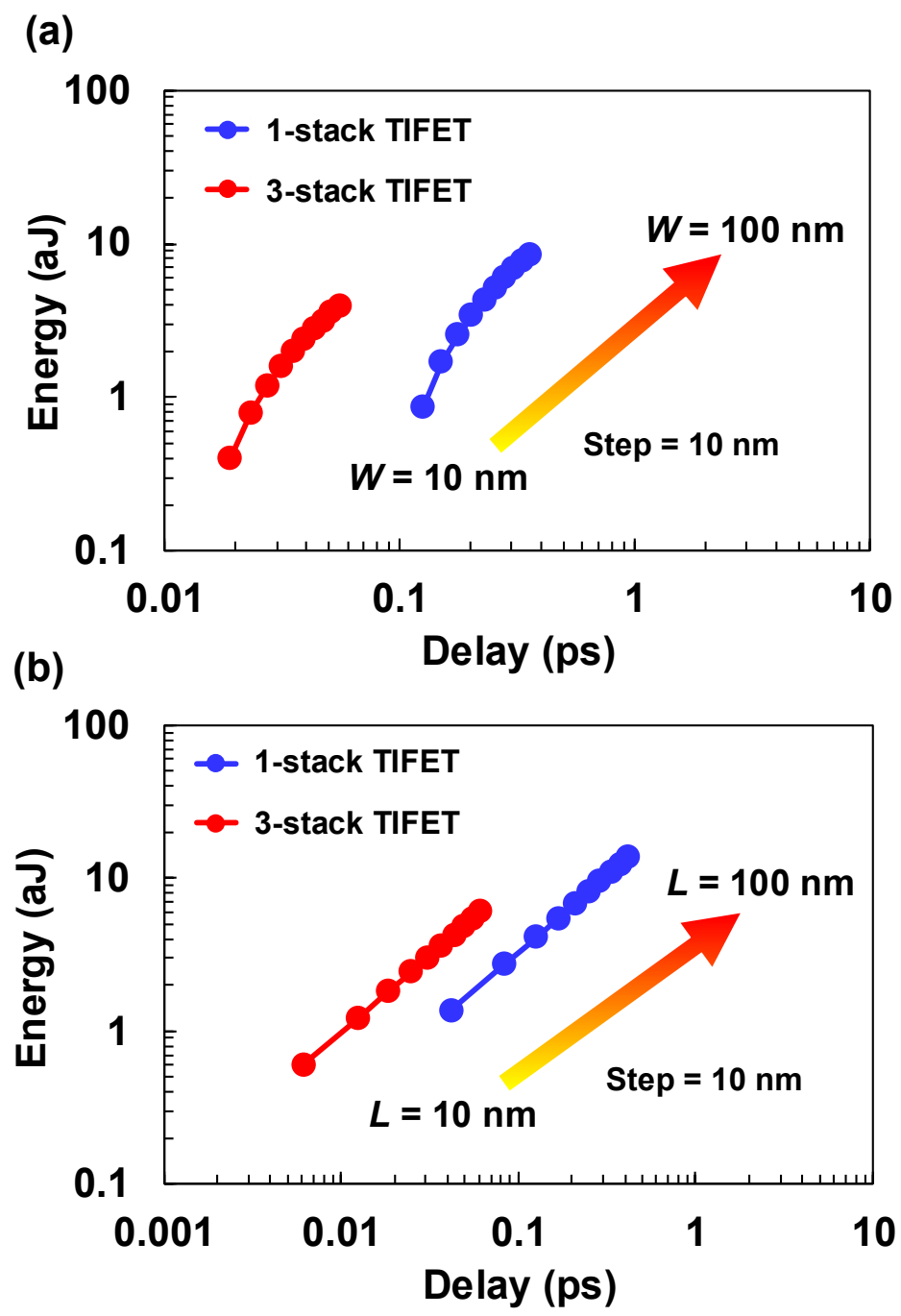

**Fig. 3.** (a) Switching energy-delay characteristics as a function of channel width, ranging from 10 to 100 nm. The channel length of both 1-stack and 3-stack TIFETs is fixed at 14 nm. (b) Switching energy-delay characteristics as a function of channel length, ranging from 10 to 100 nm. The channel width of 1-stack and 3-stack TIFETs is fixed at 30 nm. Note that $R_C$ and $C_{par}$ are not included for TIFETs.

In this study, the ON-state TIFET is modeled using the standard series-resistance equivalent circuit of conventional FETs, consisting of the source contact resistance ($R_S$), the intrinsic channel resistance ($R_{CH}$), and the drain contact resistance ($R_D$) connected in series. With symmetric contacts ( $R_S = R_D = R_C$ ), the channel resistance in the ON-state is governed by the quantum resistance of the ballistic edge channel, $R_{CH} = 1/G_0$. The voltage drops across the intrinsic channel is reduced to $V_{DD}' = V_{DD} - 2R_C I_D$ [21]. Applying Ohm's law to the intrinsic channel gives:

$$I_D = \frac{V_{DD}'}{R_{CH}} = \frac{V_{DD} - 2R_C I_D}{1/G_0}. \quad (5)$$

To incorporate the voltage drop induced by the $R_C$, $I_D$ is expressed as:

$$I_D = \frac{V_{DD}}{2R_C + 1/G_0}. \quad (6)$$

**Table 2.** Benchmarking of device performance metrics for Si-based FETs, 2D FETs, and TIFETs

| Structure | FinFET [1] | LGAA [1] | $MoS_2$ [21] | InSe [22] | 1-stack TIFET | 3-stack TIFET |
|---|---|---|---|---|---|---|
| $V_{DD}$ (V) | 0.7 | 0.65 | 0.55 | 0.5 | 0.65 | 0.65 |
| $R_{SD}$ ($\Omega \cdot \mu$m) | 271 | 257 | 246 | 124 | 124 | 124 |
| $I_D$ (μA) | 88 | 170 | 70.44 | 66.9 | 38.07 | 114.20 |
| $I_D$ (μA/μm) [a] | 874 | 787 | 1174 | 1115 | 634.46 | 634.46 |
| $L$ (nm) | 16 | 14 | 10 | 10 | 14 | 14 |
| $W$ (nm) | 5[b] | 30 | 30 | 30 | 30 | 30 |
| $W_{eff}$ (nm) [c] | 101 | 216 | 60 | 60 | 60 | 180 |
| EOT (nm) | 1 | 1 | 0.78 | 0.99 | 1 | 1 |
| $C_g$ (aF/μm) [a] | 552.27 | 483.24 | 442.70 | 317 | 101.14 | 15.59 |
| $C_{g,tot}$ (aF/μm) [a] | 1320 | 1170 | 885.40 | 634 | 202.28 | 31.17 |
| $\tau_{delay}$ (ps) | 1.060 | 0.960 | 0.415 | 0.284 | 0.207 | 0.032 |
| $E_{switching}$ (aJ) | 65.327 | 106.774 | 16.070 | 9.510 | 5.128 | 2.370 |

[a] normalized by effective channel width surrounding channels, [b] Fin width, [c] Effective channel width ($W_{eff} = 2NW$ for $N$-layer 2D FETs and TIFETs)

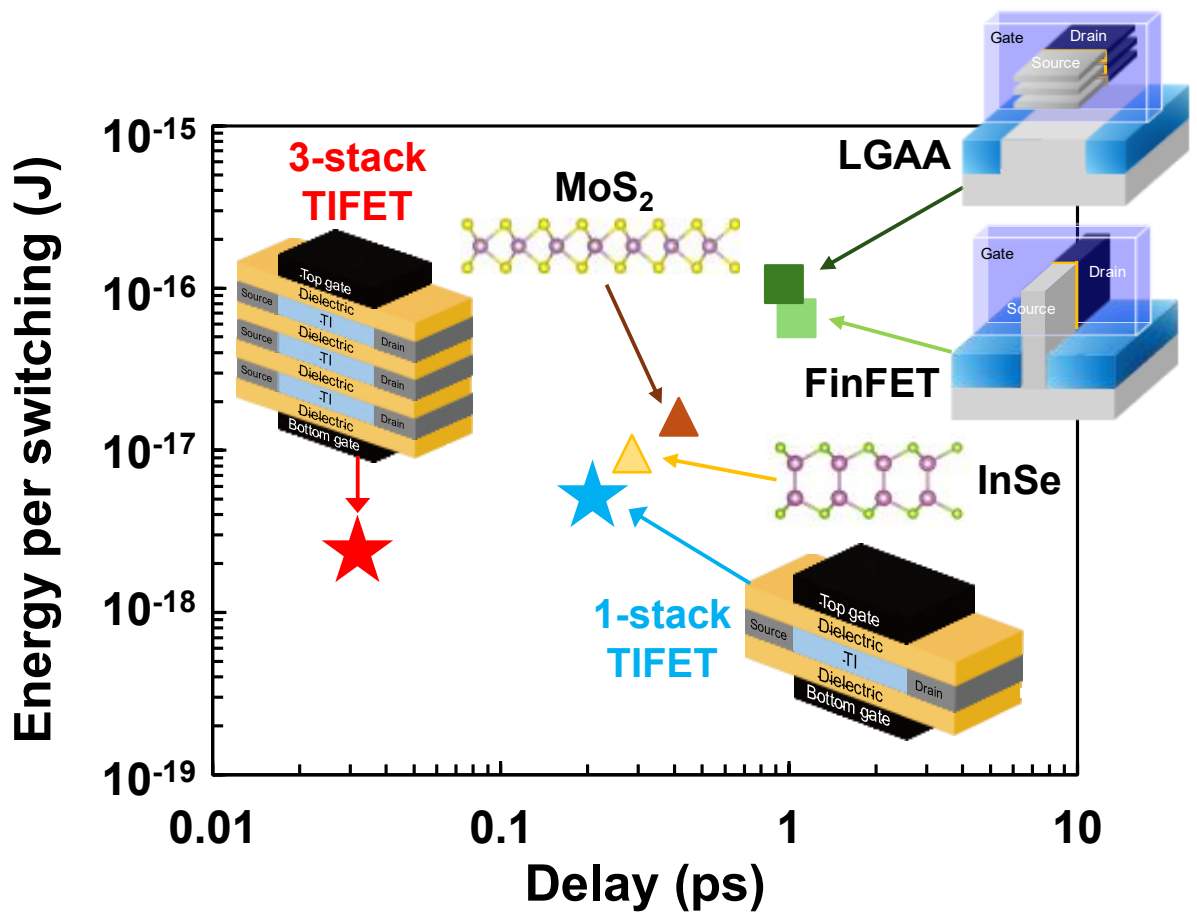

**Fig. 4.** Benchmarking of switching energy-delay characteristics of TIFETs in comparison to Si-based FETs (FinFET and LGAA) based on IRDS 2023 and 2025 targets for HP logic transistors [1] and 2D FETs ($MoS_2$ [21] and InSe [22]).

Fig. 2(a) shows the relationship between the $R_C$ and $\tau_{delay}$ of TIFETs, compared to Si-based FETs (FinFET and LGAA). The $C_{par}/C_g$ for TIFETs was set to 1.4, which approximately corresponds to the values for the FinFET and LGAA. To assess the delay performance of TIFETs, as shown in Fig. 2(a) and 2(b), their delay times were compared with those of Si-based FETs at equivalent $R_C$ and $C_{par}/C_g$ values. Under the conditions of $R_C$ = $128\,\Omega \cdot \mu$m and $C_{par}/C_g = 1.4$ consistent with IRDS targets, the delay time of both 1-stack and 3-stack TIFETs is 0.313 ps and 0.048 ps, respectively, significantly lower than that of LGAA, 0.96 ps. Given that the delay time of Si-based FETs is approximately 1 ps, the 1-stack TIFET achieves about one-third of this delay, while the 3-stack TIFET reduces the delay by approximately a factor of 20, exceeding one order of magnitude. These results demonstrate that, even when incorporating realistic contact resistance and parasitic capacitance, TIFETs can outperform conventional Si-based FETs in terms of delay time.

The switching energy-delay performance of TIFETs shown in Fig. 3 and 4 is calculated as follows: $\tau_{delay}$ is calculated using Eq. (1), and the energy per switching ($E_{switching}$) was calculated by

$$E_{switching} = C_{g,tot} \times V_{DD}^2. \quad (7)$$

In the case of Fig. 3, the effects of $R_C$ and $C_{par}$ were excluded to enable a clearer analysis of how variations in TIFET device dimensions—channel width and length—affect the switching energy-delay performance. To calculate $C_g$ under comparable conditions, the following parameters were used: the EOT was set to 1 nm to match the IRDS target [1]; the dielectric constant $\varepsilon_{ch}$ and channel thickness $t_{ch}$ were set to 5 and 0.5 nm, respectively, consistent with reported $MoS_2$ properties [30-32]. The selected values for $\varepsilon_{ch}$ and $t_{ch}$ are comparable to those of typical 2D materials (see Table 1) and are not expected to bring about any substantial alterations to the overall conclusions regarding the switching energy-delay performance of TIFETs. Additionally, $V_{DD}$ for both the 1-stack and 3-stack TIFETs was set to be 0.65 V, consistent with that of the LGAA, ensuring a fair comparison.

In conventional FETs, increasing the channel width typically reduces delay time but increases energy consumption [33]. In contrast, TIFETs exhibit the opposite behavior, where an increasing channel width results in higher energy consumption and longer delay time (see Fig. 3(a)). This is because the current of TIFETs (see Eq. (4)) is independent of channel width, as it is solely determined by the drain voltage and the quantum conductance contributed by the topological edge states. Consequently, a wider channel increases the $C_{\mathrm{g}}$, which in turn increases both delay and switching energy. As in channel width dependence, increasing the channel length in TIFETs also leads to an increase in switching energy and delay time (see Fig. 3b). As a result, an increase in channel length results in a larger $C_{\mathrm{g}}$, which leads to increased energy consumption and delay time. The results in Fig. 3 suggest that reducing the channel width and length of TIFETs is important for optimizing device performance. However, an insufficient nanoribbon width induces the finite-size effect—where finite inter-edge hopping between the two counter-propagating edge states opens an energy gap—disturbing edge states [34, 35]. While $W = 30$ nm was selected here for consistent comparison with the reference FETs, the sufficient channel width required to suppress this gap can vary with the channel material. Likewise, an excessively short channel length can induce a strong tunneling effect, degrading the OFF-state leakage [9]. Therefore, from the perspective of practical TIFET implementation, it is essential to establish an optimal channel length and width that minimizes switching energy and delay while maintaining topological switching functionally.

Fig. 4 presents a comprehensive benchmarking of the switching energy as a function of delay time of TIFETs, while Table 2 summarizes the corresponding device parameters in the TIFET model. Here, the benchmark evaluates the ON-state performance limit of TIFETs under the ideal conduction condition ($V_{\mathrm{G}} - V_{\mathrm{B}} = 0$ V). To ensure model reliability, this comparative analysis was conducted against previously validated results for Si-based FETs (IRDS targets) and experimental data for 2D FETs ($MoS_2$ and InSe). The benchmarking results reveal that both 1-stack and 3-stack TIFETs achieve switching energies more than one order of magnitude lower than those of Si-based FETs, while also exhibiting substantially reduced delay times. Furthermore, the 1-stack TIFET exhibits comparable or improved performance relative to the 2D FETs, while the 3-stack TIFET demonstrates a delay time nearly an order of magnitude lower than the 2D FETs. The benchmarked 2D FETs represent performance limits at which the detrimental effects of $R_{\mathrm{C}}$ and $C_{\mathrm{par}}$ are assumed to be minimized. Remarkably, even when considering practical $R_{\mathrm{C}}$ and $C_{\mathrm{par}}$, the projected TIFET performance remains significantly superior to these optimistic 2D FET benchmarks.

## IV. Conclusions

In conclusion, we conducted a comprehensive benchmarking study to evaluate the switching energy-delay performance limits of TIFETs in comparison with Si-based FETs and 2D-FETs. Our analysis incorporates non-ideal components, such as contact resistance and parasitic capacitance, enabling performance projections beyond ideal conditions. This highlights the potential of TIFETs to surpass Si-based and 2D FETs even under non-ideal conditions. Additionally, excluding parasitic effects reveals two key intrinsic insights: (i) vertical stacking connects gate capacitances in series while adding quantized conduction channels, uniquely enabling simultaneous improvements in delay and energy; and (ii) because the ON-current is fixed by $G_0$, this scaling is largely independent of the specific channel material, offering a universal design guideline governed primarily by gate stack and stacking number. This highlights the potential of TIFETs to surpass conventional FETs even under practical conditions. While TIFETs exhibit promising characteristics, further improvement in their on/off current ratio remains necessary. To improve the on/off performance, it is required to explore new materials that undergo topological phase transitions at low critical electric fields. Moreover, in order to enable room-temperature operation, large-bandgap topological insulators that can suppress thermal leakage need to be identified. These considerations suggest key directions for future research on TIFETs. Therefore, our benchmarking study highlights the importance of continued investigation into TIFETs as a promising candidate to overcome the scaling limits of conventional semiconductor technologies.